\documentclass[aps,english,prl,amsmath,amsfonts,amssymb,superscriptaddress,twocolumn,showpacs,citeautoscript]{revtex4-1}
\usepackage[T1]{fontenc}
\usepackage[latin9]{inputenc}
\usepackage{amsmath}
\usepackage{amssymb}
\usepackage{wasysym}
\usepackage{esint}
\usepackage{graphicx}
\usepackage{times}
\usepackage{nicefrac}
\usepackage{appendix}
\usepackage{textcase}
\usepackage{subfigure}
\usepackage{empheq}
\usepackage{color}
\usepackage{leftidx}
\usepackage{makecell}
\usepackage{stackrel}
\makeatletter
\@ifundefined{textcolor}{}
{%
 \definecolor{BLACK}{gray}{0}
 \definecolor{WHITE}{gray}{1}
 \definecolor{RED}{rgb}{1,0,0}
 \definecolor{GREEN}{rgb}{0,1,0}
 \definecolor{BLUE}{rgb}{0,0,1}
 \definecolor{CYAN}{cmyk}{1,0,0,0}
 \definecolor{MAGENTA}{cmyk}{0,1,0,0}
 \definecolor{YELLOW}{cmyk}{0,0,1,0}
}
 
\renewcommand{\Re}{\operatorname{Re}}
\renewcommand{\Im}{\operatorname{Im}}

\renewcommand{\b}{\beta}

\newcommand{\add}[1]{\if\a\b{{\color{red} #1}}\else{#1}\fi}

\newcommand{\citeasnoun}[1]{Ref.~\cite{#1}}

\renewcommand{\eqref}[1]{(\ref{eq:#1})}

\newcommand{\Eqref}[1]{Equation~\ref{eq:#1}}
\newcommand{\figref}[1]{Fig.~\ref{fig:#1}}
\newcommand{\Figref}[1]{Figure~\ref{fig:#1}}

\makeatother

\usepackage{babel}

\begin{document}

\title{Overcoming limits to near-field radiative heat transfer in
  uniform planar media through multilayer optimization}
\author{Weiliang Jin} \affiliation{Department of Electrical
  Engineering, Princeton University, Princeton, NJ 08544, USA}
\author{Riccardo Messina} \affiliation{Laboratoire Charles Coulomb,
  Universit\'{e} de Montpellier and CNRS, Montpellier, France}
\author{Alejandro W. Rodriguez} \affiliation{Department of Electrical
  Engineering, Princeton University, Princeton, NJ 08544, USA}

\begin{abstract}
  Radiative heat transfer between uniform plates is bounded by the
  narrow range and limited contribution of surface waves. Using a
  combination of analytical calculations and numerical gradient-based
  optimization, we show that such a limitation can be overcome in
  complicated multilayer geometries, allowing the scattering and
  coupling rates of slab resonances to be altered over a broad range
  of evanescent wavevectors. We conclude that while the radiative flux
  between two inhomogeneous slabs can only be weakly enhanced, the
  flux between a dipolar particle and an inhomogeneous
  slab---proportional to the local density of states---can be orders
  of magnitude larger, albeit at the expense of increased frequency
  selectivity.  A brief discussion of hyperbolic metamaterials shows
  that they provide far less enhancement than optimized inhomogeneous
  slabs.
\end{abstract}
\maketitle

\section{Introduction}

Radiative heat transfer (RHT) between two bodies separated by a
vaccuum gap and held at different temperatures is limited by
Stefan--Boltzmann's law in the far field, i.e. for gap distances $d$
much larger than the thermal wavelength $\hbar c/k_BT$ (several
microns at ambient temperature). This limitation no longer applies in
the near field, where the physics is dramatically altered by the
presence of evanenscent tunneling of photons. For instance, the
coupling of bound surface modes (e.g. phonon-polaritons in
dielectrics) can result in orders-of-magnitude larger flux
rates~\cite{basu2009review,volokitin2007near,joulain2005review,otey2014fluctuational,liu2015outlook}. Frequency-selective
bounds to RHT between arbitrarily shaped homogeneous bodies were
recently established~\cite{miller2015shape}, demonstrating that
typically considered configurations (e.g. two parallel slabs of
ordinary materials) tend to be highly suboptimal. The possibility of
increasing both far and near-field heat exchange through geometry has
attracted a remarkable amount of attention over the last several
deades due to its relevance in many technological applications, from
thermophotovoltaic (TPV) energy conversion~\cite{basu2009review} to
near-field thermal lithography~\cite{litography}.

In the far field, where the traditional, ray-optical form of
Kirchoff's law (equating emissivity and absorptivity) is
applicable~\cite{wang2011direct}, computational advances have made it
possible to exploit a variety of optimization strategies (exploiting a
variety of methods, e.g. random-walk, genetic and particle swarm
algorithms, and the Taguchi method) to realize, for instance,
selective and/or wide-angle absorbers whose emissivity can come close
to the blackbody
limits~\cite{Celanovic2004,Drevillon2007,david2007optimization,Ghebrebrhan2009,sergeant2009design,chen2010profile,chen2011polarization,Nefzaoui2012,WangSolar,Simovski2013,wangLDOS,anderson2015optimized},
and whose objective is usually that of increasing the performance of a
TPV device or solar cell. More recently, development of adjoint
optimization techniques in combination with fast numerical EM solvers
have allowed application of large-scale optimization~\cite{JensenRev}
methods capable of efficiently tackling problems involving much higher
number of degrees of freedom. For instance, these inverse-design
techniques have been exploited to enhance the far-field efficiency of
solar-cell absorbers~\cite{Ghebrebrhan2009,Ganapati}, tailor the
spectrum of incandescent sources~\cite{NatureInc}, and to increase the
functionality of photonic-crystal absorbers~\cite{Borel} and TPV
systems~\cite{Bermel2010}. Far less explored is the near field, where
the possibility of tuning and amplifying RHT has been investigated
mainly through parametric studies, i.e. shape optimization, involving
only a few degrees of freedoms. For example, several authors have
studied the role of thin films in amplifying the heat flux between
planar
objects~\cite{BiehsEJB07,ben2009near,wangF2011,boriskina2015enhancement}. Others
have focused on enhancing desirable optical properties by examining
variations with respect to Drude or Drude--Lorentz model
parameters~\cite{wang09,wangF2011,Zhao12,Nefzaoui2013}, e.g. through
doping~\cite{RousseauAPL09,BasuJH2010,Zhao12}. Various geometries such
as dielectric~\cite{LiuAPL14,fu2006nanoscale,Liu2015enhanced} and
metallic~\cite{Guerout2012grating,Dai1,Dai2,Dai3,YangPRL,MessinaarXiv,chalabi2016focused,fernandez2017enhancing}
gratings, and even finite
bodies~\cite{ramezan2016radiative,chalabi2015effect,chalabi2015near}
have been recently considered. Less restrictive inverse-design
techniques have been recently employed to improve the performances of
heat-assisted magnetic recording (HAMR)
head~\cite{Bhargava}. Relatedly, some authors have also addressed the
modulation and optimization of a closely related quantity, the
near-field electromagnetic local density of states (LDOS). For
instance,~\citeasnoun{liLDOS} performed a parametric study of the LDOS
close to a multilayer arrangement of silicon carbide and silicon thin
slabs as a function of distance and number of layers
while~\citeasnoun{benabdallahLDOS09} employed genetic algorithms to
optimize the LDOS in proximity of a multilayer binary structure
composed of alluminium and lossless dielectric layers.

In this paper, we apply adjoint, large-scale RHT optimization with the
goal of enhancing RHT in two of the most commonly studied scenarios of
two planar parallel slabs and a dipolar particle in proximity to a
slab. In particular, we relax the typical assumption of homogeneous
materials and consider instead arbitrary dielectric profiles (along
the gap direction) in planar geometries. In order to fully address the
richness of a non-uniform permittivity, we employ gradient-based
optimization over a large number of degrees of freedom (number of
layers $\gtrsim$ 1000), a regime where previously explored techniques
based on global, derivative-free optimization are bound to fail. We
demonstrate that appropriately optimized, multilayer structures can
lead to larger RHT compared to the best possible homogeneous thin
films. Our results are motivated by and extend previously derived
frequency-selective bounds for homogeneous, planar
media~\cite{miller2015shape}, quantifying the maximum flux rates that
can be achieved at any given frequency through the careful
interference or ``rate matching'' of scattered and absorbed surface
waves; such a condition can only be satisfied at a single wavevector
in uniform slabs but can be much more broadband in inhomogeneous
media. We find that with respect to uniform slabs, RHT between
inhomogeneous slabs can only be weakly enhanced, with the optimized
structures approaching the bounds.

Much larger enhancements are possible in the dipole--plate geometry,
where in principle RHT for sufficient small dipole---proportional to
the local density of states (LDOS)---can be
infinite~\cite{miller2015shape,miller2016fundamental,Maslovski16}. We
provide theoretical bounds for the flux contribution of each
wavevector in semi-infinite, uniform media and discover structures
that can approach these bounds over a broad range of wavevectors,
limited mainly by the difficult task of achieving perfect and
broadband absorption of waves close to the light line. Specifically,
we find that the optimization procedure is able to produce finite
enhancements, limited only by the finite numerical discretization
(number of layers) and sharp dielectric variations of the structures:
these can reach two orders of magnitude at mid-range $\omega d/c
\gtrsim 1$ separations but at the expense of frequency selectivity. In
the deep near field ($\omega d/c \ll 1$), on the other hand, the
condition leading to maximal LDOS is satisfied by the ideal resonant
plasmonic condition ($\Re[\epsilon] =-1$) in a uniform medium and is
therefore bounded by previously derived bounds on homogeneous
media~\cite{miller2015shape}. We remark that while the problem of
enhancing RHT between planar materials appears highly constrained and
difficult to tackle, there is much more room for improvements and
tunability when it comes to tailoring the LDOS in the vicinity of a
planar body through multilayering, which could also be of interest in
other contexts, such as in near-field
imaging~\cite{ramakrishna2003imaging,zhang2008superlenses}. 

Finally, although we only consider inhomogeneous slabs composed of
dielectric media, we argue that similar enhancements are expected in
the case of inhomogeneous, magnetically anisotropic materials. This is
exemplified, for instance, in the case of hyperbolic metamaterials
(HMM), which were recently considered in the study of radiative heat
transfer in multilayer geometries~\cite{HMMs}. In particular, focusing
on the case of uniaxial media, we show that at any separation HMMs
suffer from the same limitations of uniform thin films and thus do not
yield significant flux enhancements.

\begin{figure*}[htbp]
\centering \includegraphics[width=1.8\columnwidth]{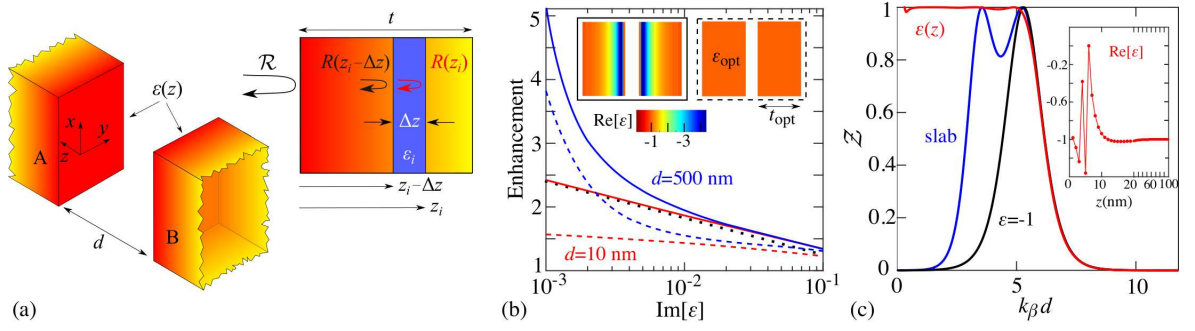}
\caption{(a) Schematic of two inhomogeneous $\varepsilon(z)$ slabs (A
  and B) separated by a vacuum gap of size $d$ along the $z$
  direction. The radiative heat transfer (RHT) rate between them
  depends on their local temperatures $T_\mathrm{A,B}$ and reflection
  coefficients $\mathcal{R}_{A,B}$. Associated with each slab is a
  coordinate system centered at the slab--vacuum interface and
  pointing away from the gap. (b) Enhancement factor comparing RHT
  between optimized inhomogeneous (solid lines) or optimized uniform
  (dashed lines) slabs against that of semi-infinite uniform plates of
  $\Re [\varepsilon]=-1$, at a fixed vacuum wavelength
  $\lambda=8~\mu$m, as a function of material loss $\Im[\varepsilon]$,
  and for two representative separations $d=10$~nm (red) and 500~nm
  (blue). The black dotted line shows the theoretical bound described
  by \eqref{bound}. (c) Transmission coefficient $\mathcal{Z}$
  corresponding to either inhomogeneous (red), uniform (blue), or
  semi-infinite (black) slabs, as a function of the dimensionless
  wavevector $k_{\beta} d$ at fixed $d=10$~nm and
  $\Im[\varepsilon]=10^{-2}$. The inset shows a typical dielectric
  profile $\Re[\varepsilon(z)]$ needed to achieve $\mathcal{Z}\approx
  1$ over a broad range of $k_\beta$. To enhance the readability, two
  different $x$-scales are used in the ranges $[0,20]\,$nm and
  $[20,100]\,$nm, and the points resulting from numerical optimization
  are connected with segments.}
\label{fig:plate}
\end{figure*}

\section{Formulation}

In what follows, we consider RHT in geometries involving slabs of
arbitrarily varying dielectric profiles $\varepsilon(z)$ along the
direction $z$ perpendicular to the slab--vacuum interfaces, depicted
in \figref{plate}(a). RHT in such a setup can be described via the
fluctuational electrodynamics framework developed by Rytov, Polder,
and van Hove (see Refs.~\cite{basu2009review,joulain2005review} and
references therein). Specifically, we extend a recently developed
formulation of this problem~\cite{messina2011pra,messina2014pra} that
expresses the flux in scenarios involving two and three uniform bodies
as a function of their reflection and transmission matrices. This
approach, together with a semi-analytic expression for the reflection
matrix of a slab of arbitrary $\varepsilon(z)$, enables gradient-based
optimizations of RHT. Since the system is time and translationally
invariant in $x$--$y$, the reflection matrix $\mathcal{R}$ is diagonal
in the frequency $\omega$, parallel wavevector
$\mathbf{k}_{\beta}=(k_x,k_y)$, and polarization $p$, and can be cast
as the solution of a differential equation, derived as
follows. Consider for each slab a local coordinate system such that
$z=0$ lies at the interface between each slab and the vacuum gap (of
size $d$) and points away from the interface. Given a slab occupying
the region $[z,t]$ (where $0<z<t$ and $t$ is the possibly infinite
thickness of the slab), let $R(z)$ be the coefficient describing the
reflection on the left side, i.e. at the interface $z$. Adding a film
of infinitesimal thickness $\Delta z$ at $z$, the reflection
coefficient of the combined system (at the $z-\Delta z$ interface) is
given by $R(z-\Delta z)=\rho(z)+\tau^2(z)R(z)/(1-R(z)\rho(z))$, where
$\rho$ and $\tau$ are the reflection and transmission coefficients of
the film, respectively. Taking the limit $\Delta z\to 0$, one obtains
the following nonlinear differential equation:
\begin{align}
  &\frac{\mathrm{d}R(z)}{\mathrm{d}z}=\frac{2ik_{zm}(z)}{1-r^2(z)}\left[r(z)\left(
    1+R^2(z)\right)-\left(1+r^2(z) \right)R(z)\right],
    \label{eq:R}
\end{align}
which, in combination with the boundary condition $R(t)=0$, describing
the absence of the slab (thus a vanishing reflection coefficient) for
$z=t$, completely specifies the reflectivity of the system. Here $r$
is the ordinary Fresnel reflection coefficient,
$k_{zm}(z)=\sqrt{\varepsilon(z)(\omega/c)^2-k_{\beta}^2}$ the
perpendicular wavevector inside the slab. Note that in the limiting
case of uniform $\varepsilon$, Eq.~\eqref{R} yields the well-known
solution
$R(z)=r[1-\mathrm{e}^{2ik_{zm}(t-z)}]/[1-r^2\mathrm{e}^{2ik_{zm}(t-z)}]$,
going to $r$ in the case of a semi-infinite slab
($t\to+\infty$). \Eqref{R} can be directly solved to obtain the
reflection coefficient of a slab of arbitrarily varying
$\varepsilon(z)$: in the case of the two slabs of \figref{plate}(a),
one would also need to specify the boundary conditions $R(+\infty)=0$
for both slabs A and B. Once the function $R(z)$ is known for each
slab, $\mathcal{R}=R(0)$ represents the reflection coefficient needed
to calculate RHT and analyze the possible enhancements arising from a
given $\varepsilon(z)$, investigated below via analytical and
optimization techniques.

We seek dielectric profiles $\varepsilon(z)$ that maximize the heat
flux $H[\mathcal{R}]$ at a given frequency. In practice, given a
choice of slab thickness, numerical evaluations require that the slab
be discretized into segments, forming a multilayer geometry. We thus
replace the function $\varepsilon(z)$ with a piecewise-constant
function that assumes values $\varepsilon_i = \varepsilon(z_i)$, with
all $\{\varepsilon_i\}_{i=1,\dots,N}$ taken as variable degrees of
freedom. Note that the size of individual layers typically needs to be
very small in order to resolve the exponential decay of evanescent
fields, with typical $N \gtrsim 100$. Furthermore, while gradient
information $\frac{\partial H}{\delta \varepsilon_i}=\frac{\partial
  H}{\partial\mathcal{R}}\frac{\partial
  \mathcal{R}}{\partial\varepsilon_i}$ is typically needed for large
$N$~\cite{Nocedal06}, $\frac{\partial
  \mathcal{R}}{\partial\varepsilon_i}$ can be straightforwardly
obtained from~\eqref{R}. Because RHT can diverge with vanishing loss
rate~\cite{miller2015shape}, (while most plasmonic material have
nonzero loss rate (unless doped with gain
media~\cite{khandekar2015giant}), we consider finite but uniform $\Im
[\varepsilon]$ thoughout the slabs, focusing only on optimizing with
respect to $\{\Re[\varepsilon_i]\}$, which leaves modifications in the
scattering rather than loss rate as the primary source of
enhancement. Since this objective function is far from
convex~\cite{boyd2004convex}, we exploit local optimization
algorithms~\cite{svanberg2002class,nocedal1980updating}.

\section{Plate--plate}\label{secplaneplane}


We first consider the scenario of two inhomogeneous, parallel slabs,
depicted in \figref{plate}(a), with both slabs A and B assumed to be
in local thermal equilibrium at temperatures $T_\text{A}$ and
$T_\text{B}$, respectively. In this case, the well-known formalism for
two slabs of uniform permittivities can be employed. The total heat
transfer per unit surface is given by~\cite{ben2009near}
$H=\int\frac{\mathrm{d}\omega}{2\pi}[\Theta(T_\text{A})-\Theta(T_\text{B})]\Phi(\omega)$,
where the monochromatic spectral component
$\Phi(\omega)=\sum_{s,p}\int\frac{\mathrm{d}k_{\beta}}{2\pi}k_{\beta}\mathcal{Z}_{s(p)}(k_{\beta},\omega)$,
with $\Theta(T)=\hbar\omega/[\exp(\hbar\omega/k_BT)-1]$ denoting the
Planck function. Here, we focus on the $p$ polarization, which
supports surface modes and hence dominates RHT at short separations
$\omega d/c \lesssim 1$. $\mathcal{Z}$ is known as the heat
transmission coefficient, whose evanescent component is given by:
\begin{equation}
  \mathcal{Z}(k_{\beta},\omega)=\frac{4\Im[\mathcal{R}_\text{A}]\Im[\mathcal{R}_\text{B}]\mathrm{e}^{-2\Im[k_z]d}}{\left|1-\mathcal{R}_\text{A}\mathcal{R}_\text{B}\mathrm{e}^{-2\Im[k_z]d}\right|^2}
\label{eq:T}
\end{equation}
The extension of \eqref{T} to the case of inhomogeneous slabs consists
in replacing the reflection operators with those from~\eqref{R}.

It is well known that energy transfer is optimal when the scattering
and absorption decay rates of the surface modes described by \eqref{T}
are equal~\cite{liu2015theory}. This corresponds to a maximum
transmissivity $\mathcal{Z}=1$, realized at
$\mathcal{R}_{\mathrm{A}}\mathcal{R}_{\mathrm{B}}^{*}=\mathrm{e}^{2\Im[k_z]d}$.
For two uniform semi-infinite slabs, such a ``rate-matching''
condition can only occur at a single $k_\beta$, depending on the
separation and loss rate~\cite{miller2015shape}, in which case
$\mathcal{Z}$ exhibits a typical Lorentzian lineshape as a function
of $k_\beta$, whose peak lies close to a typical cutoff wavevector
$k_{\mathrm{max}}$, above which $\mathcal{Z}$ is exponentially suppressed. For
small separations and loss rates and assuming operation close to the
surface plasmon resonance of a uniform slab,
i.e. $\Re[-1/\chi_\mathrm{spp}]=1/2$, such a cutoff can be
approximated by~\cite{miller2015shape}
$k_{\mathrm{max}}\approx
\frac{1}{2d}\ln[\frac{|\chi_\mathrm{spp}|^4}{4(\Im\chi_\mathrm{spp})^2}]$,
where $\chi=\varepsilon-1$ is the susceptibility of the material. This
leads to an upper bound on the RHT between uniform semi-infinite
slabs, given by
$\Phi_0=\frac{1}{2\pi
  d^2}\ln[\frac{|\chi_\mathrm{spp}|^4}{4(\Im\chi_\mathrm{spp})^2}]$~\cite{miller2015shape}.

Relaxing the assumption of uniform $\varepsilon$ allows modes of
different $k_{\beta}$ to experience different scattering and
absorption rates, potentially allowing rate-matching to not only
persist over all $k_{\beta}\lesssim k_{\mathrm{max}}$ but even beyond
$k_{\mathrm{max}}$. The latter condition, however, appears to be prohibitive. As a matter of fact, already in the case of two uniform slabs of finite thickness, the coefficient $\mathcal{Z}$ can in
principle approach 1 at arbitrarily large $k_\beta$, but only at the
expense of exponentially diverging
$\Re[\varepsilon]=-\Im[\varepsilon]\mathrm{e}^{k_{\beta}d}$, and
vanishing thickness
$t=\frac{2}{\Im[\varepsilon]k_{\beta}}\mathrm{e}^{-k_{\beta}d}$ and
bandwidths $\Delta k_\beta = k_{\beta}\mathrm{e}^{-k_{\beta}d}$,
making such an intereference effect highly impractical if at all
feasible to sustain over a wide range of $k_\beta$. We find, however,
that there exist structures that can achieve rate matching over all
$k_{\beta}\lesssim k_{\mathrm{max}}$ and therefore whose flux is
described by a larger upper bound $\tilde{\Phi}_0$, obtained by
integrating \eqref{T} with $\mathcal{Z}=1$ up to $k_\mathrm{max}$,
given by:
\begin{equation}
  \frac{\tilde{\Phi}_0}{\Phi_0}=\frac{1}{8}\ln
  \left[\frac{|\chi_{\mathrm{spp}}|^4}{4(\Im\chi_{\mathrm{spp}})^2}
    \right]+\frac{1}{2},\label{eq:bound}
\end{equation}
The ratio $\frac{\tilde{\Phi}_0}{\Phi_0}$ depends only on material
loss, increasing with decreasing loss, as shown in \figref{plate}(b)
(black dotted line). In practice, however, such an enhancement tends
to be relatively small because of the logarithmic power law and the
fact that inhomogeneity seems to barely increase $k_\mathrm{max}$,
which is instead primarily determined by the choice of loss rate and
$d$.

Next, we exploit optimization to discover inhomogeneous structures
that can achieve or approach the monochromatic bounds on
$\Phi(\omega)$ above. Although we consider the permittivities of the
two slabs to be independent degrees of freedom, we find that the
optimization always leads to a symmetric dielectric profile,
guaranteeing the rate matching condition over a wide range of
$k_{\beta}$. The inset of \figref{plate}(c) shows $\varepsilon(z)$ for
one such optimized slab, corresponding to the particular choice of
$\Im[\varepsilon]=10^{-2}$ and $d=10$~nm at the vacuum wavelength
$\lambda=8~\mu$m (frequency $\omega = 2\pi
c/\lambda\approx2.35\cdot10^{14}$~rad/s), with each dot representing
the permittivity of a $1$~nm-thick layer. The function
$\Re[\varepsilon(z)]$ shows a strong variation near the slab--vacuum
interface, approaching $-1$ away from the interface. This somewhat
un-intuitive dielectric profile leads to nearly perfect
$\mathcal{Z}=1$ for all $k_{\beta} < k_{\mathrm{max}}\approx 5/d$,
shown in \figref{plate}(c) (red solid line). In contrast, the
transmissivity of either a uniform, semi-infinite slab of
$\varepsilon=-1$ (black solid line) or a finite slab of optimal
thickness $t_\mathrm{opt}$ and permittivity $\varepsilon_\mathrm{opt}$
(blue solid line) exhibit $\mathcal{Z}\sim 1$ over a smaller range of
$k_{\beta}$. Moreover, we find that these enhancements are robust with
respect to frequency and layer thicknesses on the order of
10~nm. \Figref{plate}(b) shows the enhancement factor associated with
two different structures, optimized to maximize RHT at either
$d=10$~nm (red lines) or $d=500$~nm (blue lines), as a function of the
loss rate. We find that at small $d=10$~nm, the achievable
enhancements agree well with the predictions of \eqref{bound} while at
larger $d=500$~nm and smaller loss rates, larger enhancements are
observed; such a discrepancy is expected since the non-retardation
approximation employed in deriving \eqref{bound} underestimates
$k_\text{max}$ at mid-range separations $\omega d/c \gtrsim 1$. Even
then, the flux rates of inhomogeneous slabs (solid lines) tends to be
larger than those of uniform slabs (dashed lines).

\begin{figure*}[htbp]
  \centering
  \includegraphics[width=1.8\columnwidth]{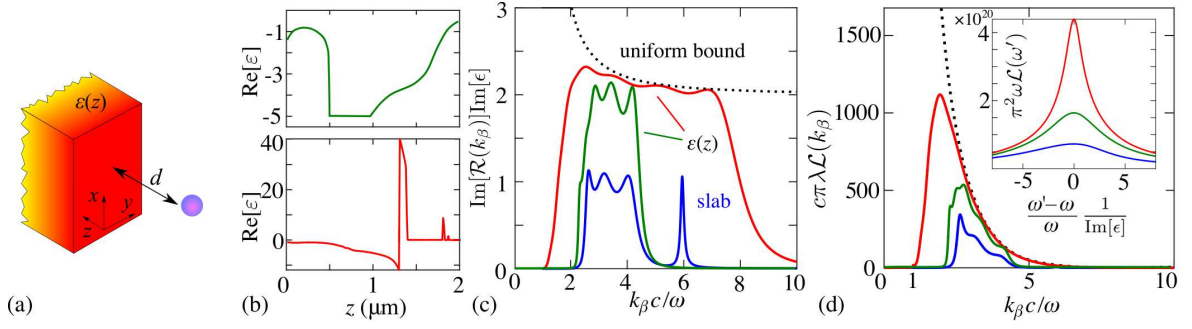}
  \caption{(a) Schematic of an inhomogeneous $\varepsilon(z)$ slab
    and a dipole separated by a vacuum gap of size $d$ along the $z$
    direction. (b) Dielectric profile $\Re[\varepsilon(z)]$ corresponding
    to inhomogeneous slabs optimized to increase RHT from a dipole a
    distance $d=1$~$\mu$m away from their $z=0$ interface, at a fixed
    vacuum wavelength $\lambda=8~\mu$m (frequency $\omega \approx
    0.785c/d$). The profiles are obtained under different constraints
    on the maximum possible permittivity $\varepsilon_\mathrm{max}
    \equiv \mathrm{max}|\!\Re \varepsilon|=\{5,40\}$ (upper and lower
    figures) but correspond to the same uniform
    $\Im[\varepsilon]=10^{-3}$.  The imaginary part of the reflection
    coefficient $\Im[\mathcal{R}(k_\beta)]$ and LDOS
    $\mathcal{L}(k_\beta)$ (in SI units) at the location of the dipole
    and at $\omega$ are plotted in (b) and (c) as a function of
    $k_{\beta}$, along with those of optimized uniform slabs (blue
    solid line). The black dotted line shows the theoretical bound
    described by \eqref{boundR}. The inset in (c) shows the
    $k_\beta$-integrated spectrum $\mathcal{L}(\omega')$ near $\omega$
    as a function of the dimensionless frequency $(\omega' -
    \omega)/\omega\Im[\varepsilon]$, indicating that contributions
    from smaller $k_\beta$ are increasingly sensitive to the
    wavelength.}
  \label{fig:sp}
\end{figure*}

While the configuration of isotropic (possibly inhomogeneous) parallel
slabs explored above yields only a small enhancement factor, stemming
primarily from the logarithmic power law and difficulty of increasing
$k_{\mathrm{max}}$, one might ask whether it is possible to further
increase $k_{\mathrm{max}}$ by exploting more exotic media,
e.g. electric and magnetic anisotropy. For the sake of simplicity, we
restrict our discussion to uniaxial media, described by diagonal
permittivity and permeability tensors given by:
\begin{equation}
  \varepsilon = \begin{pmatrix}
    &\varepsilon_{\parallel}&&\\ &&\varepsilon_{\parallel}&\\ &&&\varepsilon_{\perp}
  \end{pmatrix},\;
  \mu=
\begin{pmatrix}
  &\mu_{\parallel}&&\\
  &&\mu_{\parallel}&\\
  &&&\mu_{\perp}
\end{pmatrix}.
\end{equation}
For such media, RHT is still described by Eq.~\eqref{T}, but with a
modified expression of the perpendicular component of the wavevector
inside the medium, which reads
$k_{zm}=\sqrt{\varepsilon_{\parallel}\mu_{\parallel}k_0^2-\frac{\varepsilon_{\parallel}}{\varepsilon_{\perp}}k_{\beta}^2}$
and $\sqrt{\varepsilon_{\parallel}\mu_{\parallel}k_0^2-
  \frac{\mu_{\parallel}}{\mu_{\perp}}k_{\beta}^2}$ for the $p$ and $s$
polarizations, respectively~\cite{hu2002characteristics}. Moreover,
the corresponding Fresnel reflection coefficients have to be modified
and become
$r^p=\frac{\varepsilon_{\parallel}k_z-k_{zm}}{\varepsilon_{\parallel}k_z+k_{zm}}$
and
$r^s=\frac{\mu_{\parallel}k_z-k_{zm}}{\mu_{\parallel}k_z+k_{zm}}$~\cite{hu2002characteristics}. Since
the reflection coefficients of the two polarizations are symmetric
with respect to exchange of $\varepsilon$ and $\mu$, implying the
existence of both electric or magnetic
phonon-polaritons~\cite{francoeur2011electric}, one can focus on only
one polarization, e.g. the $p$ polarization. In the extreme near-field
regime, where the non-retarded approximation is valid, the
corresponding reflection coefficient is well approximated by
\begin{equation}
  r^p\approx
  \frac{i\varepsilon_{\parallel}/\sqrt{-\frac{\varepsilon_{\parallel}}{\varepsilon_{\perp}}}-1}{i\varepsilon_{\parallel}/\sqrt{-\frac{\varepsilon_{\parallel}}{\varepsilon_{\perp}}}+1},
\end{equation}
and is therefore equivalent to the reflectivity of an isotropic medium
of effective permittivity
$\varepsilon_{\mathrm{iso}}=i\varepsilon_{\parallel}/\sqrt{-\frac{\varepsilon_{\parallel}}{\varepsilon_{\perp}}}$.
Thus, in analogy to the uniform isotropic medium at a fixed
separation, the key to increase $k_{\mathrm{max}}$ is to reduce
$\Im[\varepsilon]$ at the surface-resonance frequency, defined by the
resonance condition $\Re[\varepsilon_{\mathrm{iso}}]=-1$, for which we
have, assuming the same loss rate,
$\Im[\varepsilon_{\parallel}]=\Im[\varepsilon_{\perp}]\equiv
\Im[\varepsilon]\ll 1$,
\begin{equation}
  \Im[\varepsilon_{\mathrm{iso}}]\approx
  \frac{1}{2}\left(\sqrt{\frac{\Re[\varepsilon_{\parallel}]}{\Re[\varepsilon_{\perp]}}}+\sqrt{\frac{\Re[\varepsilon_{\perp}]}{\Re[\varepsilon_{\parallel]}}}\,\,\right)\Im[\varepsilon]\geq\Im[\varepsilon].
\end{equation}
Thus, one concludes that the anisotropy does not allow one to decrease
losses and hence increase the cutoff wavevector $k_{\mathrm{max}}$.

\section{Dipole--slab}

In this section, we study RHT between an inhomogeneous slab and a
dipole, depicted in \figref{sp}(a). Beginning with a brief overview of
the formulation, we establish an approximate bound for RHT in the case
of semi-infinite, homogeneous bodies and exploit optimization to show
that multilayer structures can come close to approaching these limits
over a broad range of wavevectors, leading to orders-of-magnitude
larger RHT. Our work extends recent studies of near-field RHT between
dipoles and HMMs or thin films~\cite{miller2014effectiveness} to the
mid-field regime.

Consider a small sphere of radius $R\ll d$, approximated as a dipolar
particle of polarizability $\alpha$, $d$ being its distance from a
planar substrate [see \figref{sp}(a)]. In what follows, we focus on
the off-resonance regime $\alpha \mathcal{D}_{ll}\ll 1$, in which the
spectral transfer rate reads
$\Phi(\omega)=4\sum_{l=x,y,z}\Im[\alpha(\omega)]\Im[\mathcal{D}_{ll}(\omega)]$,
where $\mathcal{D}_{ll}$ denotes the Green's function of the slab at
the position of the dipole~\cite{volokitin2007near}. At short
separations $\omega d/c \lesssim 1$, the relevant tensor components of
the Green's functions are:
\begin{align}
  &\mathcal{D}_{xx}=\mathcal{D}_{yy}=\frac{i}{2}\int_{0}^{\infty}
  \mathrm{d}k_{\beta}k_zk_{\beta} \left(
  1-\mathcal{R}\mathrm{e}^{2ik_zd}
  \right),\nonumber\\ &\mathcal{D}_{zz}=i\int_{0}^{\infty}
  \mathrm{d}k_{\beta}\frac{k_{\beta}^3}{k_z}\left(1+\mathcal{R}\mathrm{e}^{2ik_zd}
  \right),
    \label{eq:G}
\end{align}
with the reflection coefficient $\mathcal{R}$ of the slab obtained
again by solving Eq.~\eqref{R}. Note that by Poynting's theorem, the
RHT rate is proportional to the LDOS at the position of the dipole,
$\mathcal{L}(\omega)=\frac{1}{2\pi^2\omega}\sum_{l=x,y,z}\Im[\mathcal{D}_{ll}(\omega)]$,
except in the regime $k_{\beta}<\omega/c$ where the latter
overestimates RHT since it also captures power radiating into the
vacuum region~\cite{oskooi2013electromagnetic}.

First, as in the plate--plate scenario, we investigate the possible
enhancements in $\Phi(\omega)$ or $\mathcal{L}(\omega)$ that can arise
from a spatially varying dielectric profile. Unlike the previous
scenario, where $\Phi[\mathcal{R}]$ was a highly non-monotonic
function of $\mathcal{R}$, here the RHT integrand is linearly
proportional to $\Im[\mathcal{R}]$. Assuming small losses
$\Im[\varepsilon]\ll 1$ and defining $k=ck_{\beta}/\omega$, a useful
figure of merit is the maximum $\Im[\mathcal{R}(k)]$ and optimal
permittivity of a uniform, semi-infinite slab at any given $k$:
\begin{align}
  &\Im[\mathcal{R}(k)]=\frac{1}{\Im[\varepsilon]}
    \frac{2+4k^2(k^2-1)+2\sqrt{1+4k^2(k^2-1)}}{(k^2-1)\sqrt{1+4k^2(k^2-1)}},\label{eq:boundR}\\
  &\Re[\varepsilon(k)]=-\frac{1+\sqrt{1+4k^2(k^2-1)}}{2(k^2-1)}.\label{eq:epR}
\end{align}
Both quantities are strongly divergent at $k=1$, suggesting that the
monochromatic LDOS $\mathcal{L}(\omega)$ of an inhomogeneous slab can
in principle be unbounded, with the main contribution to the
divergence coming from wavevectors near the light cone $k_\beta
=\omega/c$. We first observe that such a divergence would
theoretically persist for any distance $d$, since the separation
enters~\eqref{G} only as a parameter through the exponential
factors. However,~\eqref{epR} shows that at least for a uniform
medium, maximizing $\Im \mathcal{R}$ in the limit $k\to1$ requires a
perfect metal ($\varepsilon\to-\infty$), which can be shown to screen
the response at other $k$, resulting in a vanishing bandwidth $\Delta
k = 2(k-1)^2\Im \varepsilon$ and $\mathcal{L}(\omega)\rightarrow
0$. Consequently, the integrated response $\mathcal{L}(\omega)$ of a
uniform slab is finite and maximized by a finite thickness and
permittivity, which can be found numerically. More significant
improvements, however, can be gained from a spatially varying
$\varepsilon(z)$, which provides additional degrees of freedom with
which to simultaneously tune the scattering rate at different $k$,
allowing the response to approach the bounds described
by~\eqref{boundR} over wider bandwidths. Realizing such an enhancement
presents, however, both conceptual and numerical challenges: waves
approaching the light line have increasingly longer wavelength in the
$z$ direction and are thus increasingly sensitive to spatial
variations, requiring longer slabs and sharper variations in
$\varepsilon(z)$. Any numerical optimization strategy will thus
benefit only from finite enhancements coming from $k\gtrsim1$ due to
the finite number of layers needed to resolve $\varepsilon(z)$. One
should also consider that, as shown above, in the simple case of a
uniform slab the permittivity that maximizes the LDOS at $k \gg 1$
equals -1, while~~\eqref{epR} requires $\varepsilon = -\infty$ as
$k\to1$. In practice, the optimal profile results from a tradeoff
between these two conditions, since very high values of $\varepsilon$
act to screen the response from other regions of the slab. Such
distinct and challenging requirements make the optimization procedure
highly nontrivial, increasing the computational cost of RHT
calculations and slowing the convergence rate of the optimization
algorithm, which can get easily trapped in multiple local optima. To
illustrate these features, we perform separate optimizations with
different constraints on the maximum allowed permittivity
$\varepsilon_\mathrm{max}=\text{max}\{|\!\Re \varepsilon|\}$, which
limits potential enhancements coming from waves near the light line.

\begin{figure}[htbp]
  \centering \includegraphics[width=1\columnwidth]{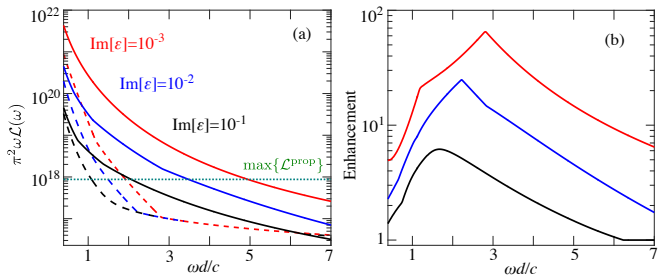}
  \caption{(a) Peak LDOS of optimized inhomogeneous slabs (solid
    lines) and optimized uniform slabs (dashed lines), as a function
    of the dimensionless separation $\omega d/c$, for multiple
    material loss rates $\Im[\varepsilon]=\{10^{-1},10^{-2},10^{-3}\}$
    (black, blue, and red, respectively). The green dotted line marks
    the largest possible LDOS in the far field, given by
    \eqref{limitff}. (b) Enhancement factor comparing the peak LDOS of
    optimized inhomogeneous and uniform slabs.}
  \label{fig:sp2}
\end{figure}

Figure~\ref{fig:sp} reports optimizations of the evanescent
contribution to $\mathcal{L}(\omega)$ at a vacuum wavelength
$\lambda=8~\mu$m and for $d=1~\mu$m. \Figref{sp}(b) shows
representative profiles $\Re[\varepsilon(z)]$ obtained under different
$\varepsilon_\mathrm{max}=\{5,40\}$ and at fixed
$\Im[\varepsilon]=10^{-3}$. Noticeably, the lower profile exhibits
rapid, subwavelength variations over small (tens to hundreds of
nanometers) regions: as discussed earlier, these are needed to
maximize $\Im[\mathcal{R}]$ near $k_\beta=\omega/c$ while avoiding
screening effects at larger $k_\beta$, with higher
$|\!\Re[\varepsilon]|$ occurring away from the interface for the same
reason. This explains why larger $\varepsilon_\mathrm{max}$ lead to
greater enhancements, illustrated in Figs.~\ref{fig:sp}(c) and (d),
which show $\Im[\mathcal{R}]$ and $\mathcal{L}$ as a function of
$k_\beta$.  The results, which are also compared against those of
uniform slabs of optimal thickness and permittivity (blue line),
reveal that inhomogeneous structures can approach the bounds described
by~\eqref{boundR} (dashed black line) over much broader range of
$k_\beta$.  Although producing a significant increase with respect to
uniform slabs, the optimization fails as $k_\beta \to \omega/c$ due to
the practical limitations discussed above. Moreover, these
enhancements will necessarily come at the expense of increased
frequency selectivity, since waves near the light line are most
sensitive to deviations in the long-range spatial pattern of the
structure, here optimized to realize a specific interference pattern
at $\omega$. This feature is apparent from the inset of
\figref{sp}(d), which shows the spectra $\mathcal{L}(\omega')$ of the
optimized uniform and inhomogeneous slabs from above (neglecting
material dispersion): namely, the contribution of lower $k$ states
becomes increasingly restricted to frequencies $\omega'\approx \omega$
as $k \to \omega/c$. (Note that the factor of $\Im \varepsilon$ in the
abscissa is there because just as in the case of a uniform medium, the
bandwidth of the Lorentzian-like spectrum is proportional to the loss
rate.)

We now explore the enhancement factor from optimized inhomogeneous
slabs for a wide range of separations $d \in [0.5,9]~\mu$m. To begin
with, \figref{sp2}(a) shows the maximum LDOS of the optimized
inhomogeneous (solid lines) and uniform (dashed lines) slabs, as a
function of $d$ and for multiple values of
$\Im[\varepsilon]=\{10^{-3},10^{-2},10^{-1}\}$ (red, blue, and black
lines respectively), with their ratio, the enhancement factor,
depicted in \figref{sp2}(b). As shown, in both situations the LDOS
increases rapidly with decreasing $\omega d/c$ and decreasing material
losses. In particular, the enhancement factor [in principle infinite
for any $d$, as suggested by~\eqref{boundR}] increases up to a maximum
value (dictated by the smallest participating, enhanced wavevector)
and then decreases, approaching 1 as $d \to \infty$. Essentially, at
large $d$, the evanescent LDOS becomes increasingly dominated by
wavevectors close to the light line (for which the optimization
procedure fails). Consequently, beyond some separation the propagating
contributions to $\mathcal{L}(\omega)$ dominate. Such finite
enhancements also imply that at small $d$, where the LDOS becomes
increasingly dominated by large $kc/\omega \gg 1$ waves, the optimal
slab is one satisfying the typical resonant condition of
$\Re[\varepsilon]=-1$ and hence the enhancement factor approaches 1.

For comparison, the green dotted line in \figref{sp2}(a) denotes the
largest achievable far-field LDOS in planar media,
\begin{equation}\label{eq:limitff}
\mathrm{max}\{\mathcal{L}^{\text{prop}}\}=\left(\frac{4}{3}+\frac{\sqrt{2}}{3}\right)\frac{\omega^2}{\pi^2c^3},
\end{equation}
derived by summing the propagative contributions under rate-matched
reflection and energy conservation, $|R_{p(s)}|\leq 1$, in which case
$|\Re[R_{p(s)}\mathrm{e}^{2ik_zd}]\,|\leq 1$. More precisely, the
limit \eqref{limitff} can be derived by observing that for the $p$
polarization,
\begin{align}
  \mathcal{L}_p^{\text{prop}}&=\frac{\omega^2}{\pi^2c^3}+\frac{\omega^2}{\pi^2c^3}\int_0^{k_0}\mathrm{d}k_{\beta}\left(\frac{k_{\beta}^3}{k_z}-k_{\beta}k_z\right)\Re[R_p\mathrm{e}^{2ik_zd}]\nonumber \\&\leq\frac{2\omega^2}{3\pi^2c^3}(1+\sqrt{2}),
\end{align}
with the maximum achieved for structures with
$\Re[R_p\mathrm{e}^{2ik_zd}]=1$ ($-1$) at
$\frac{k_{\beta}^3}{k_z}-k_{\beta}k_z>0$ ($<0$); a similar bound
applies to the $s$ polarization, leading to \eqref{limitff}. Since it
is derived under the assumption of an integrand maximized for any
$k_\beta$, $\mathrm{max}\{\mathcal{L}^{\text{prop}}\}$ provides an
upper bound that is challenging to realize. However, as shown in
\figref{sp2}, it is still smaller than the evanescent part of the LDOS
for optimized inhomogeneous slabs over a wide range of separations. In
particular, we remark that at the separations where the enhancement
factor peaks, the evanescent contribution to the LDOS of the optimized
homogeneous slabs is more than an order of magnutude larger than
$\mathrm{max}\{\mathcal{L}^{\text{prop}}\}$.

We remark that in the scenario of inhomogenous slabs, the LDOS
decreases smoothly with increasing separation and material loss, while
in the case of optimal uniform slabs, material losses become
irrelevant at distances $d\gtrsim 3c/\omega$. In order to explain this
feature, we observe that for for
$\Im[\varepsilon]=\{10^{-3},10^{-2}\}$ (red and blue lines
respectively), the first-order derivative of the peak LDOS for uniform
slabs is a discontinuous function of separation at a given
$\tilde{d}$, depending on $\Im[\varepsilon]$. This results from an
abrupt transition between two mechanisms of enhancement. In
particular, depending on the separation, the uniform slab either
maximizes the LDOS at some intermediate $k \gg 1$ through a resonant
$\Re[\varepsilon]\approx -1$ (small $d$) or near $k \approx 1$ with
$\Re[\varepsilon]\rightarrow -\infty$ (large $d$). In the latter case,
the imaginary part of the permittivity becomes irrelevant.

Since LDOS enhancements from inhomogeneous slabs prove significant at
mid-range separations, one may wonder whether similar enhancements can
be achieved in previously studied planar geometries. One such geometry
are HMMs, which consist of alternating metal and dielectric layers and
are known to exhibit hyperbolic
dispersion~\cite{guo2012broadband,BiehsAPL13}. While it has been
demonstrated that RHT between a dipole and a HMM in the deep near
field is no larger than that of an appropriately designed uniform thin
film~\cite{miller2014effectiveness}, we analyze below whether this
remains true in the mid-field regime. Consider for instance, a HMM of
period $a\ll \lambda,d$ described as an effective, uniform anisotropic
medium, with permittivities $\varepsilon_{\parallel}$
(surface-parallel) and $\varepsilon_{\perp}$ (surface-perpendicular),
having real parts $\varepsilon_1$ and $\varepsilon_2$ respectively,
and by assumption the same small imaginary part
$\Im[\varepsilon]\ll1$. The hyperbolic regions of the spectrum are
those in which $\varepsilon_1\varepsilon_2<1$, i.e. when the real
parts have opposite signs. More specifically, one typically defines
two categories of HMMs: type I with $\varepsilon_1>0$ and
$\varepsilon_2<0$, and type II with $\varepsilon_1<0$ and
$\varepsilon_2>0$. While in the extreme near-field there is no
distinction between the two types, given that only the product
$\varepsilon_1\varepsilon_2$ matters~\cite{guo2012broadband}, the two
exhibit very different behavior in the mid-field. Focusing on the
dominant, $p$ polarization ($\mathcal{R}\approx r^p$), the relevant
quantity in type-I HMMs is
\begin{equation}
\Im [\mathcal{R}_\text{I}]\approx
\frac{2\sqrt{k^2-1}\sqrt{\varepsilon_1-\frac{\varepsilon_1}{\varepsilon_2}k^2}}{\varepsilon_1(k^2-1)+1-\frac{1}{\varepsilon_2}k^2}\leq1,
\end{equation}
which is clearly far smaller than that of uniform isotropic slabs
(scaling as $1/\Im[\varepsilon]$), resulting in much smaller RHT in
the limit of small $\Im[\varepsilon] \ll 1$. The behavior of type-II
HMMs, on the other hand, varies depending on two different $k$
regimes: $k>\sqrt{\varepsilon_2}$, in which case $\Im
[\mathcal{R}_\text{II}]\approx \Im[\mathcal{R}_\text{I}]<1$, and
$1\leq k<\sqrt{\varepsilon_2}$ (requiring $\varepsilon_2>1$), in which
case the peak
\begin{equation}
  \Im[\mathcal{R}_\text{II}]\approx\frac{4\varepsilon_1\varepsilon_2(\varepsilon_2-1)}{\varepsilon_1-1+\varepsilon_2-\varepsilon_2^2}\frac{1}{\Im[\varepsilon]}
\end{equation}
occurs at
$k_m=\sqrt{\frac{(1-\varepsilon_1)\varepsilon_2}{1-\varepsilon_1\varepsilon_2}}$. Note
that similar to the case of isotropic slabs,
$\Im[\mathcal{R}_\text{II}]\rightarrow\infty$ as
$\varepsilon_1\rightarrow-\infty$ and
$\varepsilon_2\rightarrow\infty$. However, as before, such a large
dielectric constant results in a significant screening effect and thus
narrow bandwidth, whose full width at half maximum $\Delta
k\approx\frac{|\varepsilon_1-1+3\varepsilon_2(\varepsilon_2-1)|}{\sqrt{(1-\varepsilon_1\varepsilon_2)^{3}\varepsilon_2(1-\varepsilon_1)}}\Im[\varepsilon]$. In
the limit of a diverging permittivity, $\Delta k\rightarrow 0$ faster
than $\Im[\mathcal{R}_\text{II}]$ diverges, leading to vanishing
RHT. Hence, one finds once again that HMMs are in principle not better
than uniform, isotropic thin films at mid-field separations and are
therefore never ``discovered'' by the optimization method.

\section{Concluding Remarks}

We have studied optimization of frequency-selective near-field
radiative heat transfer between either two slabs or between a dipolar
particle and a slab, allowing inhomogenous dielectric profiles along
the symmetry axis of the slabs. Our approach relied on application of
large-scale, gradient-based optimization, which allows dealing with a
large number of degrees of freedom, in combination with a
simple-to-evaluate differential equation describing the reflectivity
of a planar substrate of varying $\varepsilon$. In the plate-plate
scenario, we extended previous theoretical bounds derived for
homogeneous planar structures~\cite{miller2015shape} and showed that
inhomogeneous dielectric profiles enable rate matching (perfect
absorption) of waves over a wide range of wavevectors, limited only by
material loss rates. In spite of nearly perfect coupling, we find
relatively low enhancement factors due to the logarithmic dependence
of the amplification on material losses. Nevertheless, we observe that
the plate--plate optimization is robust up to single-layer thicknesses
of the order of tens of nanometers and with respect to frequency,
making these predictions in principle experimentally feasible albeit
challenging to test, e.g. by employing molecular beam epitaxy with
vertically varying doping concentration and hence dielectric
properties.

We found, however, that much larger enhancements can be achieved in
the dipole--plate scenario, where RHT is proportional to the
LDOS. While the LDOS in this configuration is in principle unbounded,
we find that any RHT enhancement is in practice limited by a
challenging and seemingly prohibitive compromise requiring
perfect-metal behavior to increase the flux at small wavevectors
(approaching the light line) and resonant, negative permittivities
($\Re[\varepsilon]\approx-1$) needed at larger wavevectors. Our
optimization procedure was able to discover $\varepsilon$ profiles
able to partially satisfy these two criteria and hence achieve large
near-field absorption over a wide range of wavevectors, leading to
enhancement factors of up to two orders of magnitude at mid-range
separations, limited only by the finite discretization and size of the
multilayer structure as well as the existence of multiple local
minima. As explained, because the source of the enhancement is
increased absorption from waves near the light line, such increased
RHT in the dipole--plate configuration necessarily comes at the
expenses of increased frequency selectiviy and lack of robustness,
depending sensitively on the precise dielectric arrangement and choice
of wavelength. Nevertheless, our numerical experiments provide a proof
of principle that multilayer structures can overcome the
wavevector--bandwidth limitations of uniform slabs in the near field,
allowing the wavevector and spectral response of planar materials to
be tailored.


\section*{Acknowledgements} This work was supported by the National
Science Foundation under Grant no. DMR-1454836 and by the Princeton
Center for Complex Materials, a MRSEC supported by NSF Grant DMR
1420541.

\bibliographystyle{osajnl}

\end{document}